\documentclass[aps,prl,nofootinbib,showpacs,twocolumn,superscriptaddress]{revtex4-2}

\usepackage[english]{babel}
\usepackage[utf8]{inputenc}
\usepackage{amsmath}
\usepackage{amssymb}
\usepackage[caption=false]{subfig}
\usepackage{amssymb}
\usepackage{epsfig}
\usepackage{graphicx}
\usepackage{amsmath}
\usepackage{array,color}
\usepackage{natbib}

\usepackage[usenames,dvipsnames]{xcolor}
\definecolor{forestgreen}{rgb}{0.11,0.54,0.15}
\definecolor{purple}{rgb}{0.62,0.10,0.96}
\definecolor{dockerblue}{rgb}{0.11,0.56,0.98}
\definecolor{freeblue}{rgb}{0.25,0.41,0.88}

\usepackage[pdftex,plainpages=false,colorlinks=true,linkcolor=Red, citecolor=blue, urlcolor=blue]{hyperref}

\begin{document}

\title{Electrodynamics of the quantum anomalous Hall state in a magnetically doped topological insulator }

\author{Zhenisbek Tagay}
\affiliation{Department of Physics and Astronomy, The Johns Hopkins University, Baltimore, MD 21218 USA.}

\author{Hee Taek Yi}
\affiliation{Department of Physics and Astronomy, Rutgers the State University of New Jersey, Piscataway, NJ 08854 USA.}

\author{Deepti Jain}
\affiliation{Department of Physics and Astronomy, Rutgers the State University of New Jersey, Piscataway, NJ 08854 USA.}

\author{Seongshik Oh}
\affiliation{Department of Physics and Astronomy, Rutgers the State University of New Jersey, Piscataway, NJ 08854 USA.}

\author{N.~P.~Armitage}
\email{npa@jhu.edu}
\affiliation{Department of Physics and Astronomy, The Johns Hopkins University, Baltimore, MD 21218 USA.}

\date{\today}
\begin{abstract}
Magnetically doped topological insulators have been extensively studied over the past decade as a material platform to exhibit quantum anomalous Hall effect.  Most material realizations are magnetically doped and despite material advances suffer from large disorder effects.  In such systems, it is believed that magnetic disorder leads to a spatially varying Dirac mass gap and chemical potential fluctuations, and hence quantized conductance is only observed at very low temperatures.   Here, we use a recently developed high-precision time-domain terahertz (THz) polarimeter to study the low-energy electrodynamic response of Cr-doped (Bi,Sb)$_2$Te$_3$ thin films.   These films have been recently shown to exhibit a dc quantized anomalous Hall response up to T = 2 K at zero gate voltage.  We show that the real part of the THz range Hall conductance $\sigma_{xy}(\omega)$ is slightly smaller than $e^2/h$ down to T = 2 K with an unconventional decreasing dependence on frequency.  The imaginary (dissipative) part of  $\sigma_{xy}(\omega)$ is small, but increasing as a function of omega.  We connect both aspects of our data to a simple model for effective magnetic gap disorder.  Our work highlights the different effect that disorder can have on the dc vs. ac quantum anomalous Hall effect.

\end{abstract}
\maketitle

The quantum Hall (QH) effect is a phenomenon described by the emergence of dissipation-less chiral edge states that form when a large magnetic field is applied to a 2D conductor. Its striking experimental signature, quantized transverse (Hall) conductance, is quantized accurately better than 1 part in $10^{10}$ and has been used to define universal physical constants~\cite{GaAsHighaccuracy, GrapheneHighaccuracy} since it does not depend on measurement or materials details.  Haldane has emphasized that the essential aspect for quantization is time-reversal symmetry (TRS) breaking and proposed a model that had broken TRS, but at zero magnetic field~\cite{Haldane}.  Nearly fifteen years ago it was proposed that one can achieve a quantum anomalous Hall (QAH) state without external magnetic field by doping thin films of topological insulators (TIs) with ferromagnetically aligned Cr/V atoms~\cite{QAHEtheory, QAHEtheoryBST}. In this class of materials, the massless Dirac dispersion of the TIs opens up a mass gap $\Delta$ by means of spontaneous ferromagnetic order.  One expects quantized transport for temperatures $kT \ll \Delta$.

These proposals led to the experimental realization of the QAH effect first in Cr-doped~\cite{QAHEfirst} and then V-doped~\cite{QAHEsecond} (Bi,Sb)$_2$Te$_3$ thin films.  Although Hall conductance measurements in magnetically doped TIs can achieve an accuracy of  1 part per 10$^8$ of the conductance quantum $e^2/h$~\cite{OkazakiQAHE}, the temperature at which the QAH is observed (T$_{QAH}$) generally remains in sub-K region. This is despite the fact that ferromagnetism in many of these systems (as manifested by the anomalous Hall effect) is found as high as 90K~\cite{HeeTaekCBST}.  It is believed that a combination of magnetic disorder induced by magnetic dopants and charged impurities are responsible for low T$_{QAH}$~\cite{ReviewQAHE, SalehiReview, HangReview}.  In the former case, rare regions of the sample can have very small magnetic mass gaps and hence some regions of the sample can dissipate when $kT > \Delta$.  In the latter case, large chemical potential variations can lead to charge puddling; the activation energy becomes the charging energy of a puddle.  Several groups have attempted to reduce the effects of disorder by various material synthesis techniques. This includes co-doping with both Cr and V atoms~\cite{QAHEcodoping}, modulation doping~\cite{QAHEdopingmodul} and using the proximity effect~\cite{QAHEproximityYao, QAHEproximityMogi, QAHEproximityChe}. Recently Yi et al.~\cite{HeeTaekCBST} have grown trilayer heterostructure of Cr-doped (Bi,Sb)$_2$Te$_3$ (CBST) using an active magnetic capping layer scheme, which exhibits a dc QAH effect at T = 2 K without applying gating voltage. This high T$_{QAH}$ and large sample dimensions (1 cm$^2$) make this material a suitable platform to study the quantum anomalous Hall effect with various optical techniques.

The QH effect is generally observed using dc transport.  Probing the QH state with optical (ac) techniques is rather challenging, but can reveal unique information~\cite{LiangBiSe, ShimanoQHE, IkebeQHE}. In order to measure a quantized response, the probing frequencies in optical experiments have to be well below the gap, which typically constrains the relevant frequencies to sub-THz range. Moreover, the optical Hall conductance is generally measured via polarization rotation~\cite{LiangBiSe, ShimanoQHE} and rotations arising from quantized Hall conductances will be small, of order the fine-structure constant.  Therefore an optical setup has to have sub-mrad precision in polarization state detection to accurately measure the QH effect. Despite these challenges, THz measurements have been shown to have a practical advantage over dc transport when it comes to the magnetic fields needed to observe QH response.  For instance, measurements of the related quantized magnetoelectric effect via a quantized Faraday rotation in 3D thick TI films showed a quantized Faraday rotation at only 5.75 T~\cite{LiangBiSe}.  In contrast in dc transport experiments one needed to apply fields of almost 24 T in order to generate a quantized dc Hall conductance in the same film~\cite{KoiralaBiSe}.  This difference was attributed to the presence of non-chiral side states on thick films that can allow back scattering or the shorting out of the Hall voltages in a Hall bar geometry.   One expects that such side surfaces will only be localized at high magnetic fields through a combination of TRS breaking and surface roughness.   In contrast, contact-free THz measurements probe a spot in the center of TI sample and are unaffected by non-chiral edge surface states and, hence can show full quantization at much lower B. Among other aspects this has motivated us to investigate QAH state in magnetic TIs using THz spectroscopy as a potential way to minimize the effect of non-ideal sample edges on the QAH.   Moreover, the frequency dependence of the QH effect have revealed important information about its physics in the past~\cite{ShimanoQHE, IkebeQHE} and it is reasonable to expect similar experiments on the QAH effect to be similarly enlightening.

In this paper we use high-precision time-domain THz polarimetry (TDTP) to investigate the electrodynamic response of Cr-doped~\cite{QAHEfirst} (Bi,Sb)$_2$Te$_3$ (CBST) thin films at temperatures where the dc response is quantized.  Even at the lowest frequency scales we measure, the Hall conductance is smaller than the quantized value of $e^2/h$ down to T = 2 K with an unexpected decreasing dependence on frequency. We compare our data to simulations based on a simple model of a spatially disordered magnetic gap distribution and show that such disorder may be responsible for this behavior.  The extracted distribution of gap values shows consistency with those from dc transport, but different than tunneling.   This remains to be definitively understood.

We measured the polarization rotation of THz light transmitted through magnetically-doped TI films in the frequency range between 0.25 - 2.2 THz using a newly developed high-precision TDTP spectrometer~\cite{HighPrecision}.  This spectrometer is capable of measurements of the polarization state of light with a precision as good as 0.02 mrad.  After transmission through the sample, high extinction ratio wire-grid polarizers are used to split the resulting elliptically polarized THz light into $\hat{x}$ and $\hat{y}$ components, which are then captured by two separate fiber laser coupled detectors.   Data was antisymmetrized with magnetic field, i.e. $\sigma_{xy}$ at displayed field $B$ was taken to be $(\sigma_{xy}(B) - \sigma_{xy}(-B))/2  $.  Details of the setup can be found in Ref.~\cite{HighPrecision}.  The complex transmission function through a sample can be mathematically described by its Jones transfer matrix $\hat{T}$. For a material preserving $C_4$-rotation symmetry, the elements of $\hat{T}$ are constrained such that $T_{xx}=T_{yy}$ and $T_{xy}=-T_{yx}$. By performing TDTP measurements on a sample and bare substrate reference we can obtain all elements of $T$-matrix simultaneously. Note that since TDTP is a phase sensitive technique, we get complex response functions directly from experimental data without Kramers-Kronig transformations. We can then define effective transmissions for right ($r$) and left ($l$) circular bases from relation $T_{rr,ll}=T_{xx}\pm iT_{xy}$ and compute associated conductances by the usual relation for thin films on a dielectric substrate:
\begin{equation}
    \sigma_{rr,ll}(\omega)=\frac{n+1}{Z}\bigg(\frac{e^{i\Delta\phi}}{T_{rr,ll}(\omega)}-1\bigg).
\end{equation}
Here $n$ is the refractive index of substrate, $Z$ is impedance of free space ($\approx$ 377$\Omega$), and $\Delta\phi$ is a phase offset due to the thickness difference between sample and reference substrates. In the context of QH systems, it is more convenient to work with longitudinal ($\sigma_{xx}$) and transverse ($\sigma_{xy}$) conductances rather than their circular basis counterparts as we have done in Refs.~\cite{BingPbSnTe, AnaelleLSCO, BingCdAsPhonon}. Therefore, after generating $\sigma_{rr,ll}$ we convert back to Cartesian basis using relations $\sigma_{xx}=(\sigma_{rr}+\sigma_{ll})/2$ and $\sigma_{xy}=(\sigma_{rr}-\sigma_{ll})/2i$.

\begin{figure}[t!]
\centering
\includegraphics[width=0.5\textwidth]{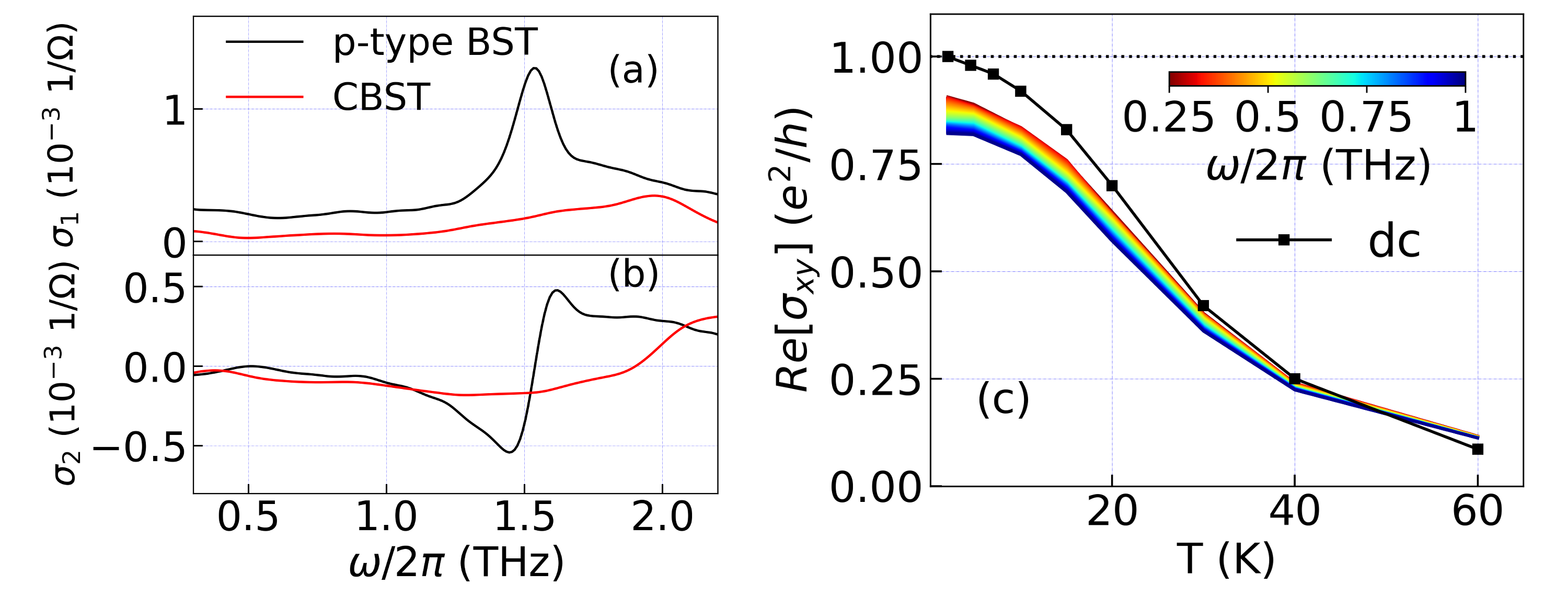} 
\label{Fig.Conductivity}
\caption{(a) Real and (b) imaginary longitudinal conductance $\sigma_{xx}(\omega)$ of p-type BST and CBST samples in zero magnetic field. (c) Temperature dependence of Re[$\sigma_{xy}$] at different frequencies. Black squares represent dc Hall transport measurements. Black dotted line indicates the quantized value of $e^2/h$. }
\end{figure}

In this work we present optical conductivity ($\sigma_{xx}$ and $\sigma_{xy}$) between 2 - 60 K in the applied magnetic field up to 6.5 T.  THz measurements were performed on multilayered heterostructures of Cr$_x$(Bi, Sb)$_{2-x}$Te$_3$ that were grown on 1 cm$^2$ Al$_2$O$_3$ substrates using  molecular beam epitaxy. The sample was composed of 4 QL Cr$_{0.1}$(Bi$_{0.143}$, Sb$_{0.857}$)$_{1.9}$Te$_3$ layer sandwiched in between 
two 3 QL Cr$_{0.48}$(Bi$_{0.13}$, Sb$_{0.87}$)$_{1.52}$Te$_3$ layers. An $a$-CrO$_x$ capping layer (0.45 nm thick) was deposited on top of the trilayer heterostucture to enhance the QAH effect and minimize its degradation effects. Sample growth details and dc transport measurements are presented in Ref.~\cite{HeeTaekCBST} (sample \#2). 

In Figs.~1 (a)-(b) we present the complex longitudinal conductance ($\sigma_{xx}=\sigma_1+i\sigma_2$) of CBST as a function of frequency $\omega$ at zero applied magnetic field.  We compare it to the optical conductance of $p$-type BST sample without magnetic dopants, which shows a sharp phonon at 1.55 THz.   This latter sample has a charge density of $1.2\times10^{12}$ cm$^{-2}$ and a Fermi energy of 30 meV below the Dirac point of the TI surface states.   Stoichiometric compounds from (Bi,Sb)-dichalcogenides family are typically characterized by two features in the THz range in zero field.  At frequencies of order 1.5 - 1.9 THz, but which depend on the particular composition, this prominent $\alpha$-mode phonon is found~\cite{DipanjanSbTe, RolandoBiSe, RolandoBiSeAging, SalehiStability, KoiralaBiSe}.  It corresponds to the mutual sliding motion of the chalcogenides layers with respect to the Bi layers.  In systems with appreciable free charge density (either surface or bulk) a prominent Drude peak is found at lower frequencies.  Its area is a measure of the charge density.  These general expectations are borne out in the $p$-type BST sample.   In contrast, for the magnetically doped CBST one can see only the vestige of the phonon as a very broad absorption over the range 1.2 - 2.2 THz.  From the lack of low frequency Drude peak, one can also see that the carrier density in this sample is very low which is an essential ingredient for observing quantized Hall response in topological insulators~\cite{LiangBiSe, KoiralaBiSe}.   The broad phonon remnant is a consequence of the extreme disorder present in the CBST heterostructure; the non-stoichiometric chemical composition of individual layers can induce significant structural disorder. In such a scenario, multiple phonon resonances of individual stoichiometric compounds can indeed smear the phonon into a single broadened mode~\cite{LiangBiInSe} similar to the wide peak observed in $\sigma_1(\omega)$ of CBST. In the case of CBST, the energy scales of the phonon modes in constituent Sb$_2$Te$_3$ and Bi$_2$Te$_3$ compounds are similar to each other and lie in the 1.5 - 1.7 THz range~\cite{DipanjanSbTe, BiTe}. Therefore, broadening of the phonon towards higher frequencies is mostly due to the substantial doping with much lighter Cr atoms.   Although the magnitude of the broadening is substantial, we note that it appears similarly broad at similar dopings in our previous study on In substitution into Bi$_2$Se$_3$~\cite{LiangBiInSe}.

In Fig.~1(c) we plot the anomalous Hall conductance (Re[$\sigma_{xy}$]) for a number of different frequencies in the 0.25 - 1 THz frequency range as a function of temperature  and compare them with dc Hall transport measurements from Ref.~\cite{HeeTaekCBST}. At each temperature, magnetic field of 1 T was first applied to induce spin polarization in the sample and then anomalous Hall conductance measurements were performed at zero field. As one can see, Re[$\sigma_{xy}$] in the THz regime doesn't reach the quantized value of $e^2/h$ down to the lowest temperature of 2 K.   At higher temperatures, the difference between the dc and THz gradually vanishes and at T = 60 K optical Hall response becomes slightly larger than in dc regime. The (non-quantized) anomalous Hall effect in this sample is reported to persist up to T = 90 K in dc Hall transport measurements~\cite{HeeTaekCBST}.  This is consistent with THz data.

To further investigate the dynamic response of the anomalous Hall state, we plot the complex $\sigma_{xy}(\omega)$ measured at different temperatures in Figs.~2 (a)-(c). For each temperature we show $\sigma_{xy}(\omega)$ both when a high magnetic field is applied and the anomalous Hall conductance after the field is turned off. The figures clearly show that Re$[\sigma_{xy}(\omega)]$ in THz range doesn't reach the quantized value of $e^2/h$, both with and without applied magnetic field.  This is remarkable as it shows that a lack of local full spin polarization is not the largest factor determining degraded quantization.   As one can see, high-field data remains largely unaffected up to T = 10 K, whereas the anomalous Hall conductance slightly decreases with temperature. Most notably however, for all temperatures Re$[\sigma_{xy}(\omega)]$ {\it decreases} with frequency which is in contradiction with the  general theoretical expectations for QH or QAH a system with a clean gap~\cite{ShimanoSrRuO,IkebeQHE,Okamuro,ahn2022theory} at frequencies below the gap.  Quite generally one expects a quadratic and increasing dependence on frequency.  Similar behavior was observed in previous works on other CBST-based heterostructures~\cite{MogiHalfQAHE, OkadaQAHE}. We will elaborate below how this unusual $\omega$-dependence can be obtained for a material with a disordered magnetic bandgap. The full temperature dependence of Re$[\sigma_{xy}(\omega)]$ at B = 0 T is depicted in Fig.~2(d). The magnitude of anomalous Hall conductance gradually decreases with temperature as expected, whereas the $\omega$-dependence stays qualitatively the same as at the lowest temperatures.

\begin{figure}[t!]
\centering
\includegraphics[width=0.5\textwidth]{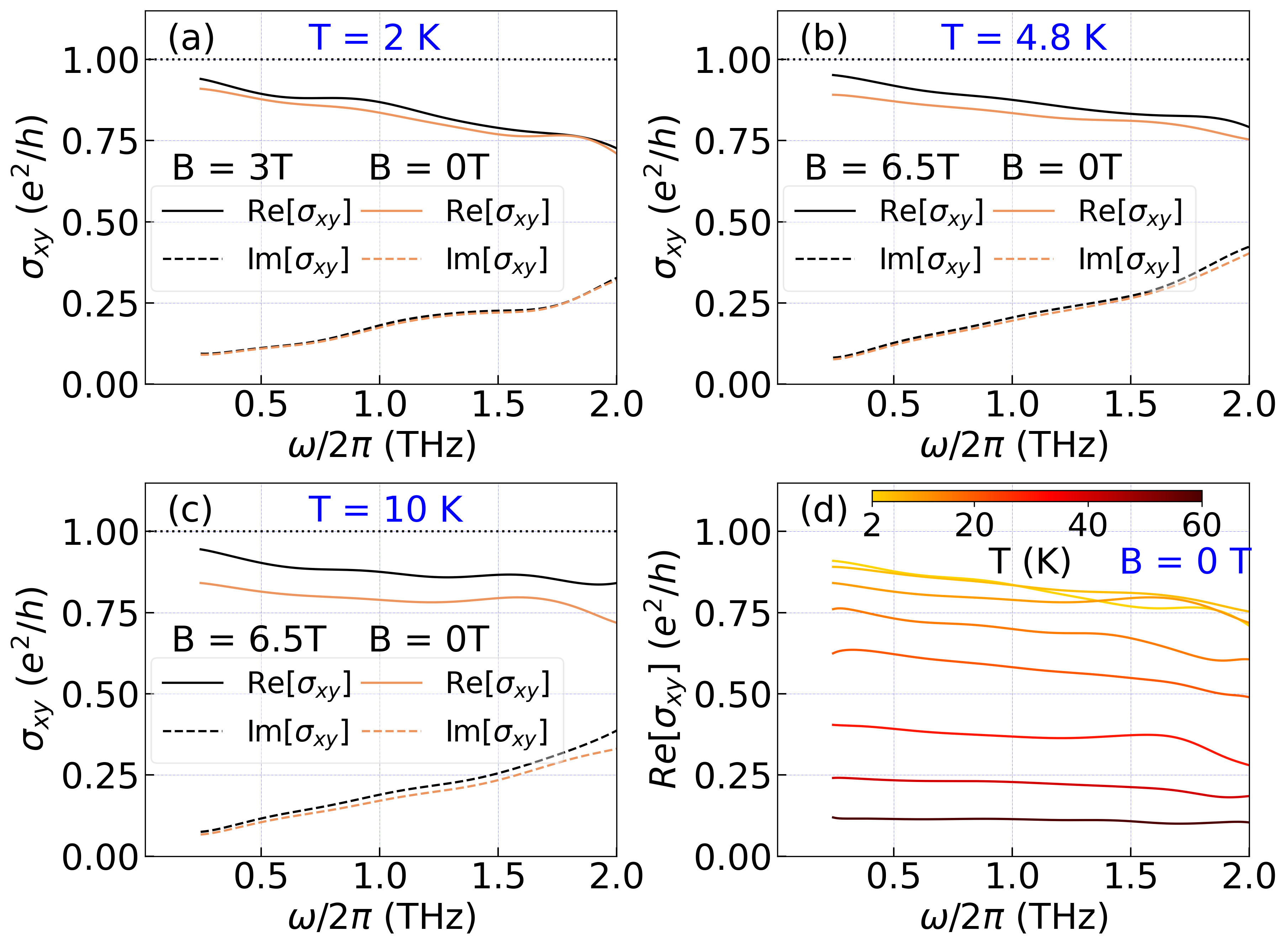} 
\label{Fig.freqdependence}
\caption{Complex Hall conductance $\sigma_{xy}(\omega)$ at (a) T = 2 K, (b) T = 4.8 K, and (c) T = 10 K. Solid and dashed lines indicate real and imaginary $\sigma_{xy}(\omega)$, respectively. Dotted lines represent the conductance quantum $e^2/h$. (d) Anomalous (B = 0 T) Hall conductance Re[$\sigma_{xy}(\omega)$] at different temperatures.}
\end{figure}

\begin{figure*}[t]
\centering
\includegraphics[width=0.85\textwidth]{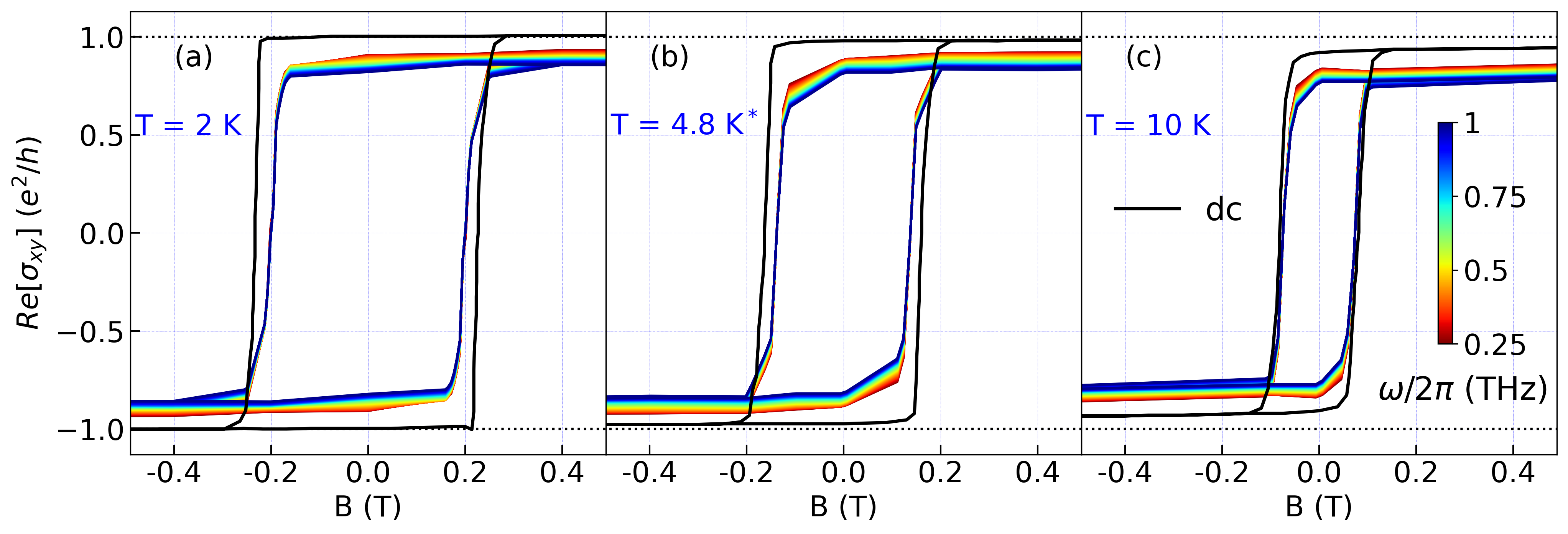} 
\label{Fig.hysteresis}
\caption{Hysteresis loops of Hall conductance Re[$\sigma_{xy}$] for different frequencies at (a) T = 2 K, (b) T = 4.8 K, and (c) T = 10 K. Solid black lines indicate dc Hall transport measurements. Dotted lines represent the conductance quantum $e^2/h$. $^*$dc data in (b) was taken at T = 4.2 K. }
\end{figure*}

In Fig.~3 we present Re$[\sigma_{xy}]$ hysteresis loops in 0.25 - 1 THz range at different temperatures. dc Hall transport measurements on the same sample from Ref.~\cite{HeeTaekCBST} are also shown for comparison. Although there is a qualitative match between the dc and THz hysteresis loop shapes, their corresponding widths and heights are noticeably different. As we mentioned before, the anomalous Hall conductance at THz frequencies does not quite reach the quantized value down to T = 2 K and is equal to 0.91$e^2/h$ at $\omega/2\pi$=0.25 THz. Corresponding values for T = 4.8 K and T = 10 K are 0.89$e^2/h$ and $0.84 e^2/h$, respectively. These numbers are all noteably lower than corresponding dc data. Moreover, we note that $\sigma_{xy}$ in THz range starts to flip its sign at smaller magnetic field than in dc measurements.   This unusual effect will be investigated in future publications.

Band structures of magnetic topological insulators can be described by a 2D massive Dirac Hamiltonian. Quite generally (in both in QH and QAH), one finds that Re[$\sigma_{xy}$] will increase quadratically as a function of probing frequency when $\omega \ll \Delta$ for material systems with a clean magnetic gap~\cite{ShimanoSrRuO, Okamuro}. However, we have found that for CBST heterostructures Re[$\sigma_{xy}$] decreases with $\omega$ for the entire frequency range between 0.25 - 2.2 THz. This indicates that models based on a clean Dirac gap cannot explain unusual behavior in THz data and disordered gap distributions need to be considered instead. It was previously reported that inhomogeneous positioning and clustering of Cr-atoms can lead to spacial fluctuations in bandgap across the surface of thin film~\cite{YoungImaging, GapDisorder, SalehiReview, HangReview, ReviewQAHE}. This disorder reduces the smallest excitation gap of the system, leading to significantly smaller value of T$_{QAH}$ as compared to Curie temperature and the spectroscopic gap values. Although $\sigma_{xy}$ in the dc regime remains quantized at temperatures below T$_{QAH}$, the Hall conductance at finite frequencies might be affected if considerable ``local'' gaps are smaller than $\omega$.  To model this possibility, we simulate  $\sigma_{xy}(\omega)$ for different bandgap distributions using a simple conductivity model that is a $\delta$-function at $\Delta$ (weighted by the magnetic gap energy $\Delta$) for the dissipative part and $\frac{e^2}{h} \frac{1}{1 - \nu^2 / \Delta^2} $ for the non-dissipative Hall response (where $\nu$ = $\omega$/2$\pi$).   Although the precise form of this function follows from the model we used, quite generally Kramers-Kronig considerations determine that for frequencies well below the scale of dissipative processes that the Hall conductivity should go quadratically {\it increasing} in frequency with a coefficient set by the inverse magnetic gap~\cite{ArmitageKK}, and there should be a negative contribution to $\sigma_{xy}$ above this scale.  Explicit calculations back up these general considerations~\cite{ShimanoSrRuO,IkebeQHE,Okamuro,ahn2022theory}.

\begin{figure}[b!]
\centering
\includegraphics[width=0.5\textwidth]{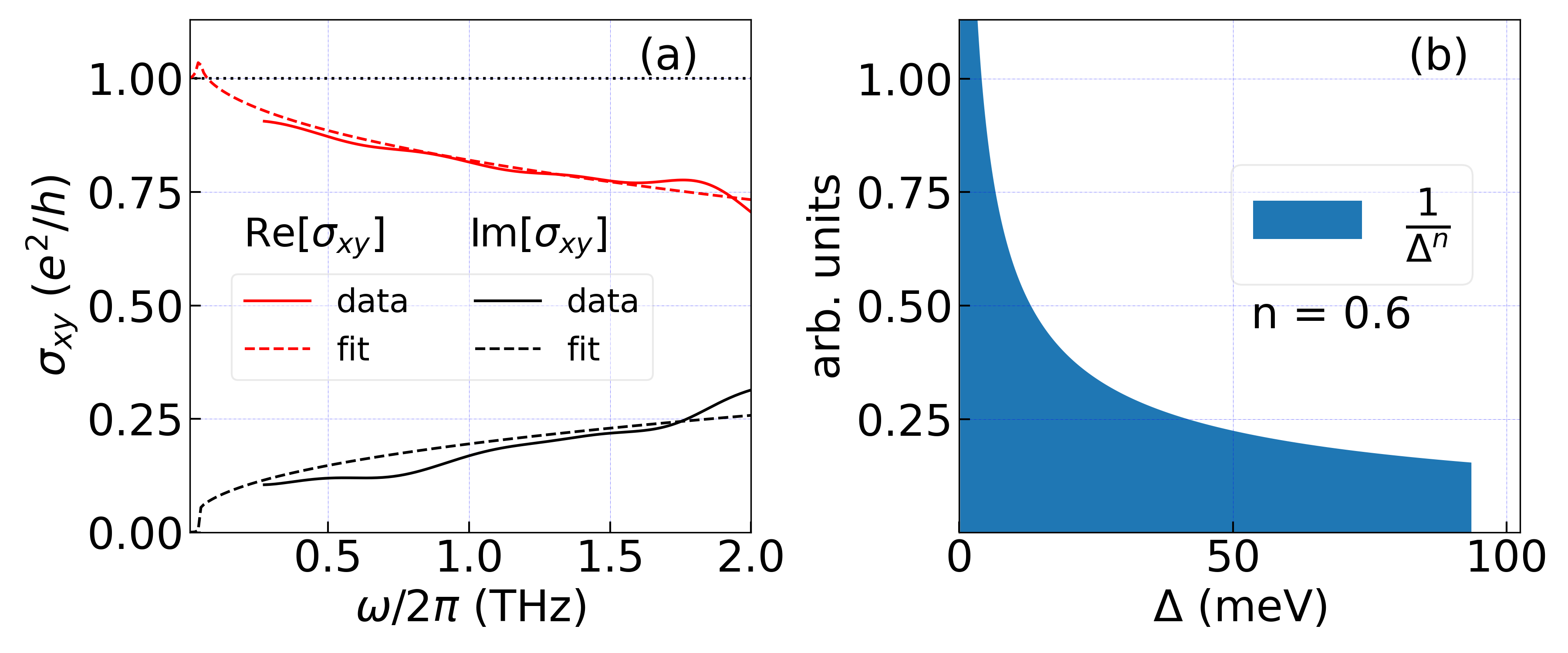} 
\label{Fig.simulation}
\caption{(a) Comparison between measured complex $\sigma_{xy}(\omega)$ at T = 2 K and B = 0 T (solid lines) and numerical simulations (dashed lines). (b) Magnetic gap distribution for the case $\Delta_{min}$ = 2 K used in the simulation}
\end{figure}

In Fig.~4(a) we show a comparison between the measured complex $\sigma_{xy}(\omega)$ at T = 2 K and B = 0 T  and a numerical simulations based on this model for Hall conductivity with a distribution of gap values. We find that a distribution with power law dependence ($1/\Delta^n$) fits real and imaginary parts simultaneously over the entire measured spectral range.  The exact value of $n$ is sensitive to our choice of lower cut-off value for gap distribution ($\Delta_{min}$). Assuming that $\Delta_{min}$ is on the order of few Kelvins at which dc transport data is still quantized, the value for $n$ can be estimated as 0.6 $\pm$ 0.1, whereas the resulting upper cut-off value ($\Delta_{max}$) varies between 70 - 130 meV. In Fig.~4(b) is the arrived distribution of gap values based on the simulations for the particular case when $\Delta_{min}$  = 2 K (0.17 meV). It is notable that this distribution of gap values is very different from that extracted from tunneling in similar samples.  There one finds a $\Delta$ distribution peaked at 30 meV and almost no regions with gaps less than 7 meV ~\cite{GapDisorder}.   This dichotomy also exists with the activated gap determined from transport.   It has been proposed that this difference originates in large fluctuations of the gap edges due to the long-range nature of the Coulomb potential of charged impurities and that at the lowest temperatures conduction occurs through activated hopping or Efros-Shklovskii variable-range hopping between electron and hole puddles created by the disorder~\cite{huang2022conductivity}.  In such a model the activation energy is set by the charging energy of a puddle.   Generalizing this picture to the ac THz case would be an interesting area for future investigation.

We have investigated the electrodynamic response of a magnetically doped TI in THz range. A broadened phonon resonance observed in longitudinal conductance of CBST indicates the presence of significant structural disorder due to non-stoichiometric chemical composition of individual layers. Unlike full quantization of $\sigma_{xy}$(B=0) observed via dc measurements, the THz range anomalous Hall conductance is systematically less than the quantized value even at our lowest frequencies. We also found $\sigma_{xy}$ exhibited an unusual decreasing frequency dependence which is in contradiction with theoretical expectation for a system with clean magnetic gap. We showed that we can simulate this $\omega$-dependence with a simple model incorporating magnetic gap disorder, however the relevant gap sizes are much smaller than those determined spectroscopically.  Our hope is that our measurements will motivate theory that will extend the models of large chemical potential fluctuations and magnetic gap disorder to ac and optical effects.

This work at JHU and Rutgers was supported by the ARO MURI ``Implementation of axion electrodynamics in topological films and device" W911NF2020166. The work at Rutgers is additionally supported by ARO grant W911NF2010108. The instrumentation development at JHU that made these measurements possible was supported by the Gordon and Betty Moore Foundation EPiQS Initiative Grant GBMF-9454.

\bibliography{main} 		
\bibliographystyle{apsrev4-1} 		


\end{document}